\documentclass[aps,prd,reprint,balancelastpage,nofootinbib,amsmath,amssymb,superscriptaddress]{revtex4-1}
\usepackage[
colorlinks=true,        
citecolor=blue,         
linkcolor=blue,         
urlcolor=blue           
]{hyperref}             
 \usepackage{graphicx}
\usepackage{bm}         
\usepackage{amsmath}

\newcommand{\nc}{\newcommand*} 
\nc{\Eq}[1]{Eq.~\eqref{#1}}     
\nc{\Fig}[1]{Fig.~\ref{#1}}     
\nc{\Table}[1]{Table~\ref{#1}}  
\nc{\Sec}[1]{Sec.~\ref{#1}}     
\nc{\red}[1]{\textcolor{red}{#1}}
\nc{\dt}{\delta}
\nc{\bt}{\mathbf{t}}
\nc{\eg}{\textit{e.g.~}}
\nc{\bn}[1]{\dt\bm{t}_{\text{#1}}}
\nc{\be}{\bm{\epsilon}}
\nc{\BF}{\mathcal{BF}}
\nc{\yr}{\mathrm{yr}}

\def\e{\begin{equation}}
\def\q{\end{equation}}

\def \kmsMpc {\mathrm{km~s^{-1}~Mpc^{-1}}}
\def \omm {\mathrm{\Omega_{m0}}}

\begin{document}
	
\title{Testing redshift variation of the   X-ray and   ultraviolet luminosity relations of quasars}
\author{Jiayi Wu}
\affiliation{Department of Physics and Synergetic Innovation Center for Quantum Effects and Applications, Hunan Normal University, Changsha, Hunan 410081, China}

\author{Yang Liu}
\affiliation{Department of Physics and Synergetic Innovation Center for Quantum Effects and Applications, Hunan Normal University, Changsha, Hunan 410081, China}

\author{Hongwei Yu}
\email{hwyu@hunnu.edu.cn}
\affiliation{Department of Physics and Synergetic Innovation Center for Quantum Effects and Applications, Hunan Normal University, Changsha, Hunan 410081, China}
\affiliation{Institute of Interdisciplinary Studies, Hunan Normal University, Changsha, Hunan 410081, China}

\author{Puxun Wu}
\email{pxwu@hunnu.edu.cn}
\affiliation{Department of Physics and Synergetic Innovation Center for Quantum Effects and Applications, Hunan Normal University, Changsha, Hunan 410081, China}
\affiliation{Institute of Interdisciplinary Studies, Hunan Normal University, Changsha, Hunan 410081, China}


\begin{abstract}
 Quasars serve as important cosmological probes and constructing accurate luminosity relations for them is essential for their use in cosmology. If the coefficients of quasar's luminosity relation vary with redshift, it could introduce biases into cosmological constraints derived from quasars. In this paper, we conduct a detailed analysis of the redshift variation in the X-ray luminosity and ultraviolet (UV) luminosity ($L_\mathrm{X}$-$L_\mathrm{UV}$) relations of quasars. For the standard $L_\mathrm{X}$-$L_\mathrm{UV}$ relation, we find that the relation coefficients   exhibit a strong and linear correlation with redshift, which is not attributable to the selection effect.  Additionally, we examine two three-dimensional, redshift-evolving $L_\mathrm{X}$-$L_\mathrm{UV}$ relations and find that  the inclusion of a redshift-dependent term does not eliminate the impact of redshift evolution, as the relation coefficients continue to evolve with redshift.  Finally, we construct  a new $L_\mathrm{X}$-$L_\mathrm{UV}$ relation  in which the redshift evolution of the relation coefficients is nearly eliminated.  Calibrating the luminosity relations using Hubble parameter measurements, we demonstrate that quasars utilizing our new relation yield effective constraints on cosmological parameters that are consistent with results from Planck CMB data, unlike constraints derived from the standard relation.


\end{abstract}

\maketitle
\section{Introduction}
Quasars (quasi-stellar objects) are  extremely luminous and persistent sources powered by gas spiraling at high velocity into supermassive black holes.  Their immense luminosity, often thousands of times greater than that of the Milky Way, make them visible across vast cosmological distances.  The maximum observed redshift of quasars can reach $z>7$\cite{Mortlock:2011,Banados:2017,Lyke2020,Wang2021} and to date,  more than half a million quasars  have been identified. If quasars can serve as  reliable cosmological probes, they offer valuable tool for filling the redshift desert in observational data. Consequently, they have been widely utilized to investigate the nature of dark energy, to explore the potential origins of the Hubble constant ($H_0$) tension,  to test the cosmic distance duality relation \cite{Zheng2020}, and so on. 

In quasar cosmology, establishing a robust luminosity relation is essential for determining quasar distances.  Several  empirical relations have been proposed, including  the anticorrelation between ultraviolet (UV) emission lines and luminosity~\cite{Baldwin1977,Osmer1999}, the luminosity-mass relation in super-Eddington accreting quasars~\cite{Wang2014}, the relation between luminosity and X-ray variability~\cite{Franca2014},  the radius-luminosity relationship~\cite{Watson2011,Melia2014,Eser2015}, the redshift-angular size  relation~\cite{Paragi1999,Chen2003,Cao2017,Cao2020a,Cao2020b,Ryan2019}, and the nonlinear relation between the X-ray luminosity ($L_\mathrm{X}$) and the UV luminosity ($L_\mathrm{UV}$)~\cite{Tananbaum1979,  Zamorani1981, Risaliti2015,Avni1986}. The $L_\mathrm{X}$-$L_\mathrm{UV}$ relation has gained particular prominence in constructing the Hubble diagram of quasars,  extending to redshifts as high as $z\sim 7.5$~\cite{Risaliti2019, Lusso2016, Lusso2017, Lusso2020, Lusso2019, Lian2021, Hu2022, Khadka2020a, Khadka2020b}.  From the Hubble diagram, the cosmological distance can be derived. Initially, the results fit the  cosmological constant plus the cold dark matter  ($\Lambda$CDM) model well and match the distances inferred from type Ia supernova (SN Ia)~\cite{Risaliti2015}. However, several groups found recently that  the distance modulus/redshift relation of quasars derived from  the $L_\mathrm{X}$-$L_\mathrm{UV}$ relation has an apparent deviation from the prediction of the $\Lambda$CDM model~\cite{Risaliti2019,Khadka2021,Li2021}.  This deviation raises concerns about the accuracy of the $L_\mathrm{X}$-$L_\mathrm{UV}$ relation  and suggests that it may evolve with  redshift~\cite{Li2022}. 

The potential redshift evolution of the 
 $L_\mathrm{X}$-$L_\mathrm{UV}$ relation has been explored in prior studies~\cite{Wang2022,Li2024}, though these analyses remain somewhat limited. For example, Wang et al \cite{Wang2022}  divided a quasar sample into low-redshift ($z<1.4$) and high-redshift $z>1.4$ subsets and found that the relation coefficients for the two subsamples differ by more than
   $2\sigma$. Similarly, Li et al.~\cite{Li2024} split a quasar sample with $z<2.261$ into two parts with almost the same data number  ($\langle z \rangle=1.16$),  found similar deviations between the relation coefficients.  While it is now recognized that the relation coefficients vary across redshift ranges, their precise evolutionary behavior remains unclear.
   
Recently, a three-dimensional and redshift evolutionary $L_\mathrm{X}$-$L_\mathrm{UV}$ relation has been constructed in~\cite{Dainotti2022}, by  considering a redshift-dependent correction to the luminosities of quasars. Additionally, Wang et al.~\cite{Wang2022} applied the statistical tool copula~\cite{Nelsen} to  constructed another three-dimensional and redshift evolutionary $L_\mathrm{X}$-$L_\mathrm{UV}$ relation.   Subsequent analyses have indicated ~\cite{Wang2024, Zhang2024} that the copula-based relation outperforms both the standard $L_\mathrm{X}$-$L_\mathrm{UV}$ relation  and the relation given in~\citep{Dainotti2022}. 
 However, whether these  generalized  relations fully eliminate   redshift variations in the  relation coefficients requires further scrutiny. 

Thus,  in this paper, we systematically   test the redshift dependence of both two- and three-dimensional $L_\mathrm{X}$-$L_\mathrm{UV}$ relations.   The paper is organized as follows. In \Sec{SIGW}, we examine the redshift variation of the standard $L_\mathrm{X}$-$L_\mathrm{UV}$ relation. In \Sec{Sec3}, we investigate whether the selection effect results in the redshift variation of the $L_\mathrm{X}$-$L_\mathrm{UV}$ relation.  In \Sec{Sec4}, we analyze two three-dimensional $L_\mathrm{X}$-$L_\mathrm{UV}$ relations. We proposes a new  $L_\mathrm{X}$-$L_\mathrm{UV}$ relation in \Sec{Sec5}. Finally, in \Sec{Sec6}, we summarize our findings.

\section{The $L_\mathrm{X}$-$L_\mathrm{UV}$ relation}\label{SIGW}

The standard $L_\mathrm{X}$-$L_\mathrm{UV}$ relation describes   a non-linear relationship between the X-ray luminosity and the UV luminosity   of quasars~\citep{Tananbaum1979,  Zamorani1981, Risaliti2015,Avni1986}, and it   takes the form
\begin{eqnarray}\label{eq.1}
	\log L_\mathrm{X}=\beta+\gamma \log L_\mathrm{UV},
\end{eqnarray}
where $L_\mathrm{X}$ and $L_\mathrm{UV}$ are the luminosities (in $\mathrm{erg~s}^{-1}~\mathrm{Hz}^{-1}$) at 2~keV and 2500 $\mathring{A}$, respectively, and  $\beta$ and $\gamma$ are two relation coefficients. Converting luminosity to flux, we can obtain 
\e
\log F_\mathrm{X}=\beta+\gamma \log F_\mathrm{UV}+(\gamma-1) \log (4\pi d_L^2),
\q
where $ F_\mathrm{X}=\frac{ L_\mathrm{X}}{4\pi d_L^2}$ and  $ F_\mathrm{UV}=\frac{ L_\mathrm{UV}}{4\pi d_L^2}$ are the fluxes of the X-ray and UV, respectively, and $d_L$ is the luminosity distance, which depends on cosmological models. 

Thus, given a cosmological model, the values of relation coefficients and the intrinsic dispersion  of quasars can be determined from the  observed quasar data. In this paper, we choose the cosmological model to be a  spatially flat $\Lambda$CDM model with $H_0=70~\kmsMpc$ and $\omm=0.3$, where $\omm$ is the present dimensionless matter density parameter. 
 
 The data sample of  quasars used in our analysis consists of 2421 X-ray and UV flux measurements, spanning a redshift range from $z = 0.009$ to $7.541$~\cite{Lusso2020}. These sources were  carefully selected   from an initial sample of 21,785 data points based on several criteria for the X-ray and ultraviolet bands, including signal-to-noise ratio, host galaxy contamination in the ultraviolet band, X-ray absorption, and Eddington bias.  These criteria were chosen to minimize observational biases. Using the maximum likelihood estimation method~\cite{DAgostini}, we obtain $\beta=6.298\pm0.228$ and $\gamma=0.665\pm0.007$ as well as the intrinsic dispersion $\delta=0.230\pm0.004$ from the  2421quasar data points.

To test whether the $L_\mathrm{X}$-$L_\mathrm{UV}$ relation varies with  cosmological redshift, we separate the  2421 quasars into four groups based on their redshift and calculate the mean redshift for each grup. The four groups, which have almost equal data numbers and cover the  redshift range from low  to high values,  are as follows:
{\small
\begin{itemize}
\item Group 1 -- 606 quasars, $0.009\leq z <0.844$, $\langle z \rangle=0.586$
\item Group 2 -- 605 quasars, $0.844\leq z <1.298$, $\langle z \rangle=1.064$
\item Group 3 -- 605 quasars, $1.298\leq z<1.892$, $\langle z \rangle=1.578$
\item Group 4 -- 605 quasars, $1.892\leq z\leq7.541$, $\langle z \rangle=2.561$
\end{itemize}}

\begin{figure}[tbp]
\centering
\includegraphics[width=0.47\textwidth]{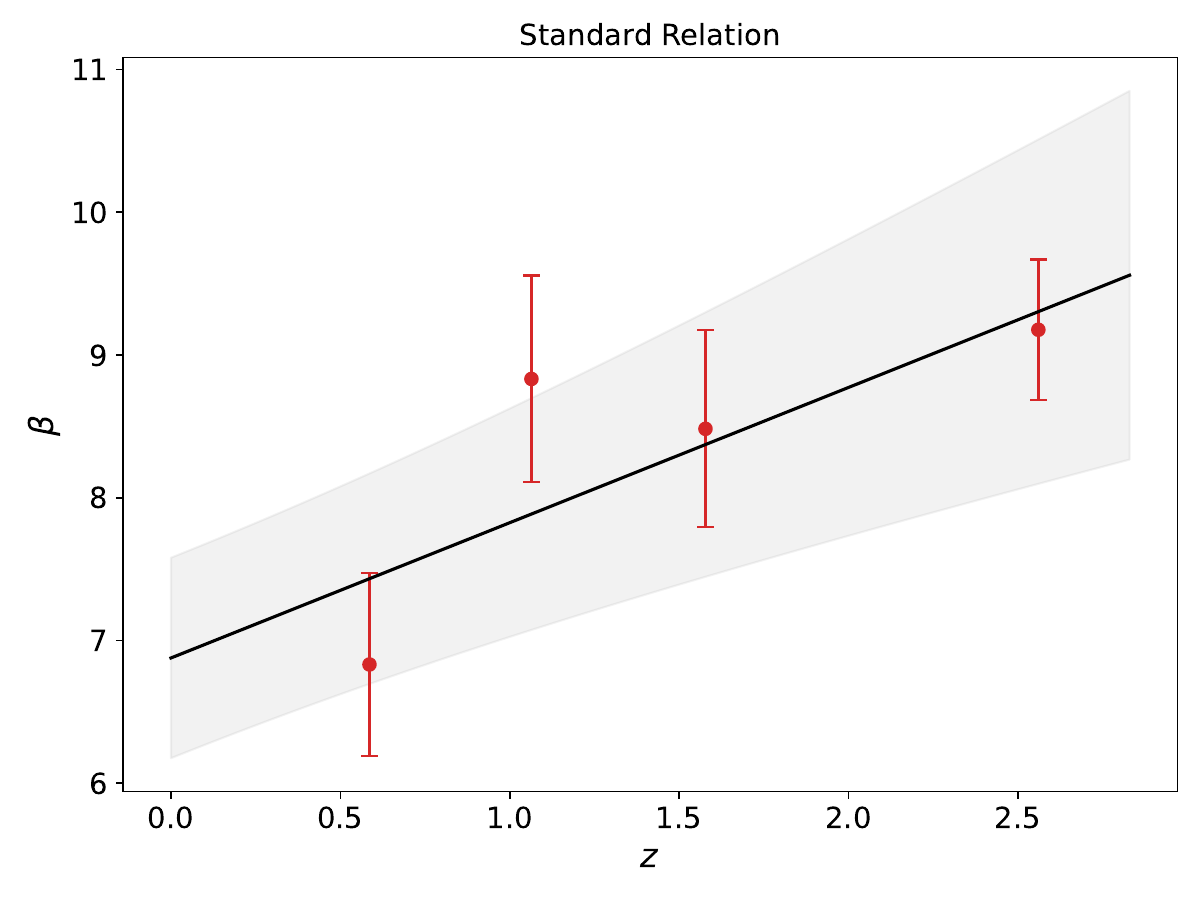}
\includegraphics[width=0.48\textwidth]{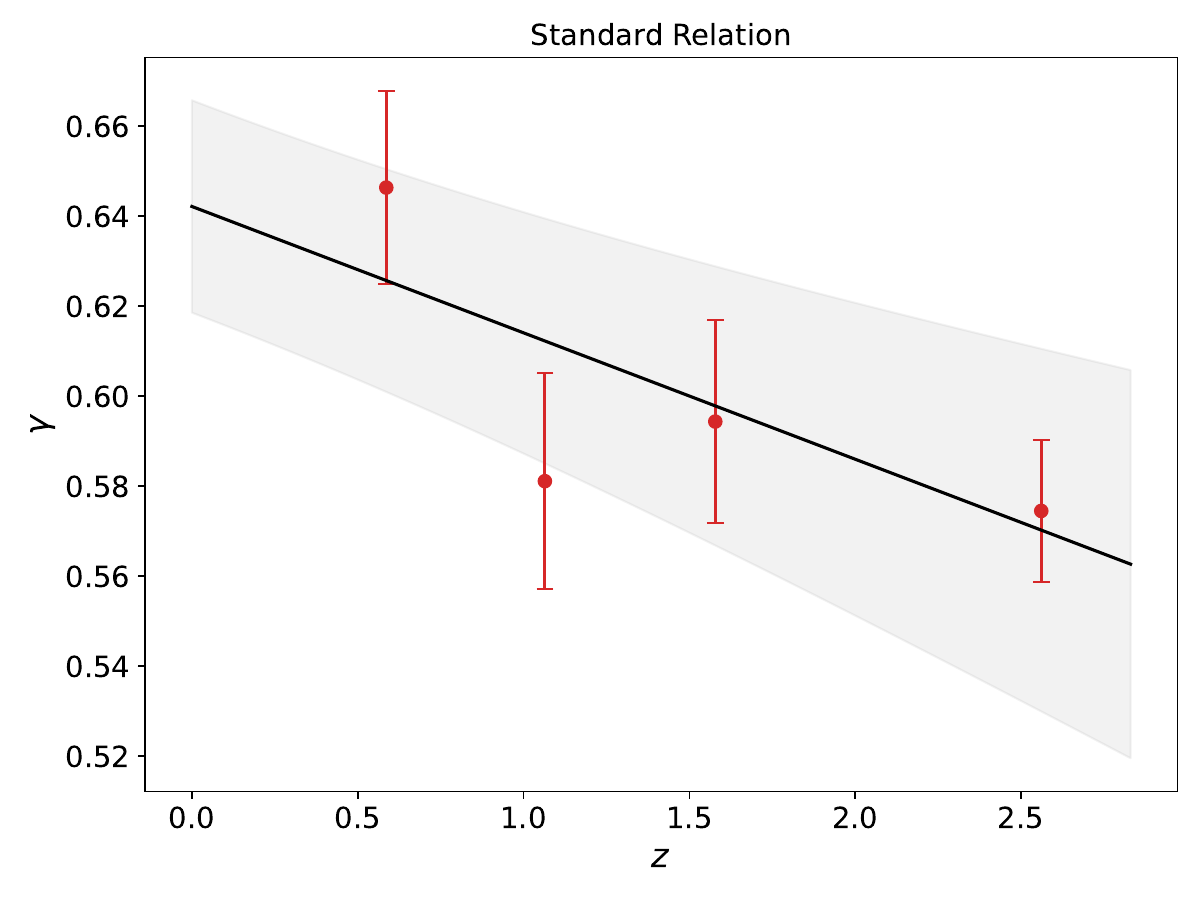}
\caption{\label{Fig1} Values of $\beta$ and $\gamma$ against the mean redshift of quasars. Each data point with 1$\sigma$ error bars represents a group of quasars (1, 2, 3, and 4). The solid line represents  a   linear fit.}
\end{figure}

We  then compare the data in each group with Eq.~(\ref{eq.1}), and then check if the values of $\beta$ and $\gamma$ evolve with  redshift. 
Utilizing the maximum likelihood estimation method~\cite{DAgostini} to fit each group of quasars,  we find that ($\beta=6.831\pm0.640$, $\gamma=0.646\pm0.021$) for Group 1, ($\beta=8.832\pm0.724$, $\gamma=0.581\pm0.024$) for Group 2, ($\beta=8.482\pm0.691$, $\gamma=0.594\pm0.022$) for Group 3, and ($\beta=9.177\pm0.492$, $\gamma=0.574\pm0.016$) for Group 4. The corresponding results are shown in Fig.~\ref{Fig1}. It is easy to see that as $\langle z \rangle$ increase,  $\beta$ increases almost  linearly, while $\gamma$ decreases linearly.  This means that $\beta$ and $\gamma$ are strongly correlated/anti-correlated with $\langle z \rangle$. We used the linear function to fit $\beta$-$\langle z \rangle$ and $\gamma$-$\langle z \rangle$, and find that 
\begin{eqnarray}\label{eq3}
	\beta &= (6.877 \pm 0.702) + (0.948 \pm 0.383) \times z 
\end{eqnarray}
and
\begin{eqnarray}\label{eq4}
	\gamma &= (0.642 \pm 0.024) + (-0.028 \pm 0.013) \times z.
\end{eqnarray}
These results show that the parameters $\beta$ and $\gamma$ tend to evolve with redshift, since the slopes of  these relationships deviate from zero by more than $2\sigma$ confidence level (CL).

To minimize the impact of grouping strategies on the redshift evolution trends  in the luminosity relation coefficients, 
we also divide the quasar sample into four groups with equal redshift intervals:
{\small
\begin{itemize}
	\item Group 1 -- 460 quasars, $0.009\leq0.750$, $\langle z \rangle=0.520$
	\item Group 2 -- 959 quasars, $0.750\leq1.500$, $\langle z \rangle=1.095$
	\item Group 3 -- 644 quasars, $1.500\leq2.250$, $\langle z \rangle=1.817$
	\item Group 4 -- 263 quasars, $2.250\leq3.000$, $\langle z \rangle=2.529$
\end{itemize}}
Here we exclude quasars with redshifts exceeding 3 since only 95 data points  fall within this range.
We use the maximum likelihood estimation method to fit each group of data separately and obtain ($\beta=6.450\pm0.718$, $\gamma=0.659\pm0.024$) for Group 1, ($\beta=8.165\pm0.532$, $\gamma=0.603\pm0.018$) for Group 2, ($\beta=10.104\pm0.595$, $\gamma=0.542\pm0.019$) for Group 3, ($\beta=8.703\pm0.837$, $\gamma=0.590\pm0.027$) for Group 4.
The corresponding linear relations for $\beta-\langle z \rangle$ and $\gamma-\langle z \rangle$ are:
\begin{eqnarray}
	\beta = (6.391\pm0.754) + (1.482\pm0.492)\times z
\end{eqnarray}
and 
\begin{eqnarray}
	\gamma = (0.657\pm0.026) + (-0.044\pm0.017)\times z.
\end{eqnarray}
Clearly, the slops of these relations remain significantly different from zero at  more than 2$\sigma$ CL, consistent with the results shown in Eqs.~(\ref{eq3}, \ref{eq4}). Therefore,  employing different grouping strategies does not change the redshift evolution trends of the parameters $\beta$ and $\gamma$.

\section{Selection effect}\label{Sec3}

In order to check whether the evolutions of $\beta$ and $\gamma$ with redshift originate from the selection effect  in the measurement process, we follow the method in \cite{Li2007} to generate a sample of quasars by using the Monte-Carlo simulations. 
Since the observed quasars provide only X-ray and UV fluxes (\textit{i.e.}, $\log F_\mathrm{UV}$ and $\log F_\mathrm{X}$), we first calculate their luminosities as $\log L_\mathrm{UV}(\log L_\mathrm{X})=\log F_\mathrm{UV}(\log F_\mathrm{X})+\log 4\pi d_L(z)^2$ in the framework of  a flat $\Lambda$CDM model with $H_0=70~\kmsMpc$ and $\omm=0.3$.  Using the real sample of 2421 quasars, we then model the distributions of redshift, $\log L_\mathrm{UV}$, and  measurement  errors $\sigma_\mathrm{UV}$ of quasars, respectively.  It has previously been demonstrated that the quasar redshift distribution follows a Gamma distribution~\cite{Wang2022},  described by $f(z) = \frac{b^a z^{a-1} e^{-b z}}{\Gamma(a)}$ with  $a=3.036$ and $b=2.098$, we therefore use this Gamma distribution to generate the redshifts for the mock quasars.  Additionally, the distribution  of $\sigma_\mathrm{UV}$ is modeled by a Lognormal distribution, given by $f(\sigma_\mathrm{UV}) = \frac{1}{\sqrt{2\pi} s\: \sigma_\mathrm{UV}} \exp\left\{ -\frac{(\log \sigma_\mathrm{UV} - \mu)^2}{2 s^2} \right\}$ with  $\mu=-4.342$ and $s=0.566$ derived directly from the real quasar data  employing the Python package {\it scipy.stats}.  
The distribution of $\log L_\mathrm{UV}$ follows a Gaussian form with a mean value of $30.428$ and a standard deviation of $0.658$, also derived  from the real data.   Although  observational limitations (e.g., flux detection thresholds or Eddington bias) could introduce deviations from the Gaussian assumptions for the UV luminosity,  we have verified the validity of this assumption through a statistical distribution test, specifically  the Cram\'{e}r-von Mises test~\cite{Cramer, Mises}. 
Then, we sample the mock datasets of ($z$, $\log L_\mathrm{UV}$, $\sigma_\mathrm{UV}$)  from these distributions. Note that  we assume no correlations between the distributions, meaning they are mutually independent.
 We then introduce a UV luminosity cutoff by removing mock data with $\log L_\mathrm{UV}<\log L_\mathrm{UV,min}$, where $\log L_\mathrm{UV,min}=\log F_\mathrm{UV,min}+\log 4\pi d_L(z)^2$ with $\log F_\mathrm{UV,min}=-28.759$ being the minimum observable flux. 
For each  triplet ($z$, $\log L_\mathrm{UV}$, $\sigma_\mathrm{UV}$), we sample $\log L_\mathrm{X}$ from the probability distribution function: $f(\log L_\mathrm{X})=\frac{1}{\sqrt{2\pi}\sigma}\exp\left\{-\frac{(\log L_\mathrm{X}-\beta-\gamma \log L_\mathrm{UV})^2}{2\sigma^2}\right\}$, where the intrinsic dispersion $\sigma$, the coefficients $\beta$ and $\gamma$ are fixed to  $\sigma = 0.230$, $\beta = 6.298$, and $\gamma=0.665$, which are obtained in the previous section.
The mock observed error $\sigma_\mathrm{X}$ is also sampled by assuming a Lognormal distribution with  mean values and standard deviations of $-3.159$ and $0.769$, respectively, from the real data.  After  imposing  an X-ray luminosity limit, analogous to the UV cutoff,  by using $\log F_\mathrm{X,min}=-32.706$,  we ultimately obtain the simulated quasar data  set ($z$, $\log L_\mathrm{UV}$, $\sigma_\mathrm{UV}$, $\log L_\mathrm{X}$, $\sigma_\mathrm{X}$).

\begin{figure}[htbp]
\centering
\includegraphics[width=0.45\textwidth]{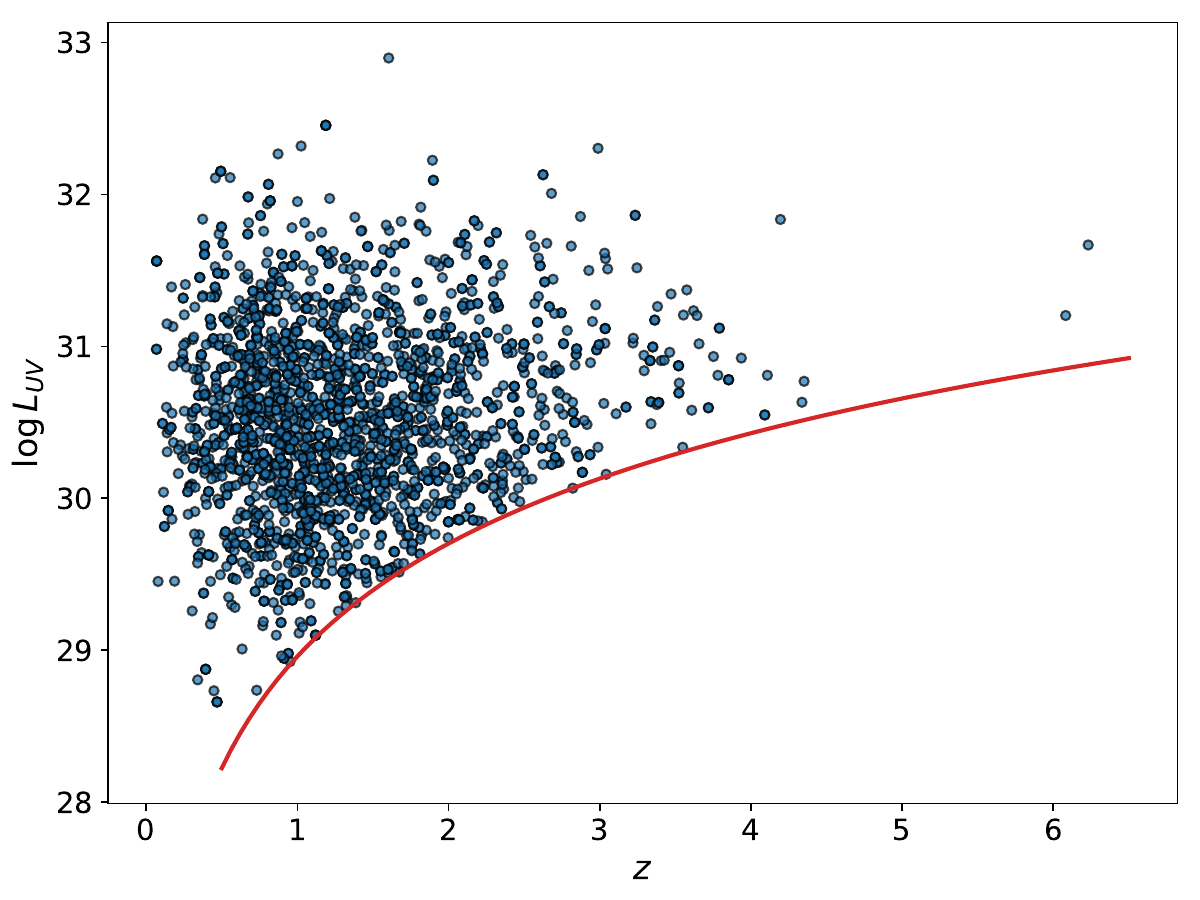}
\includegraphics[width=0.45\textwidth]{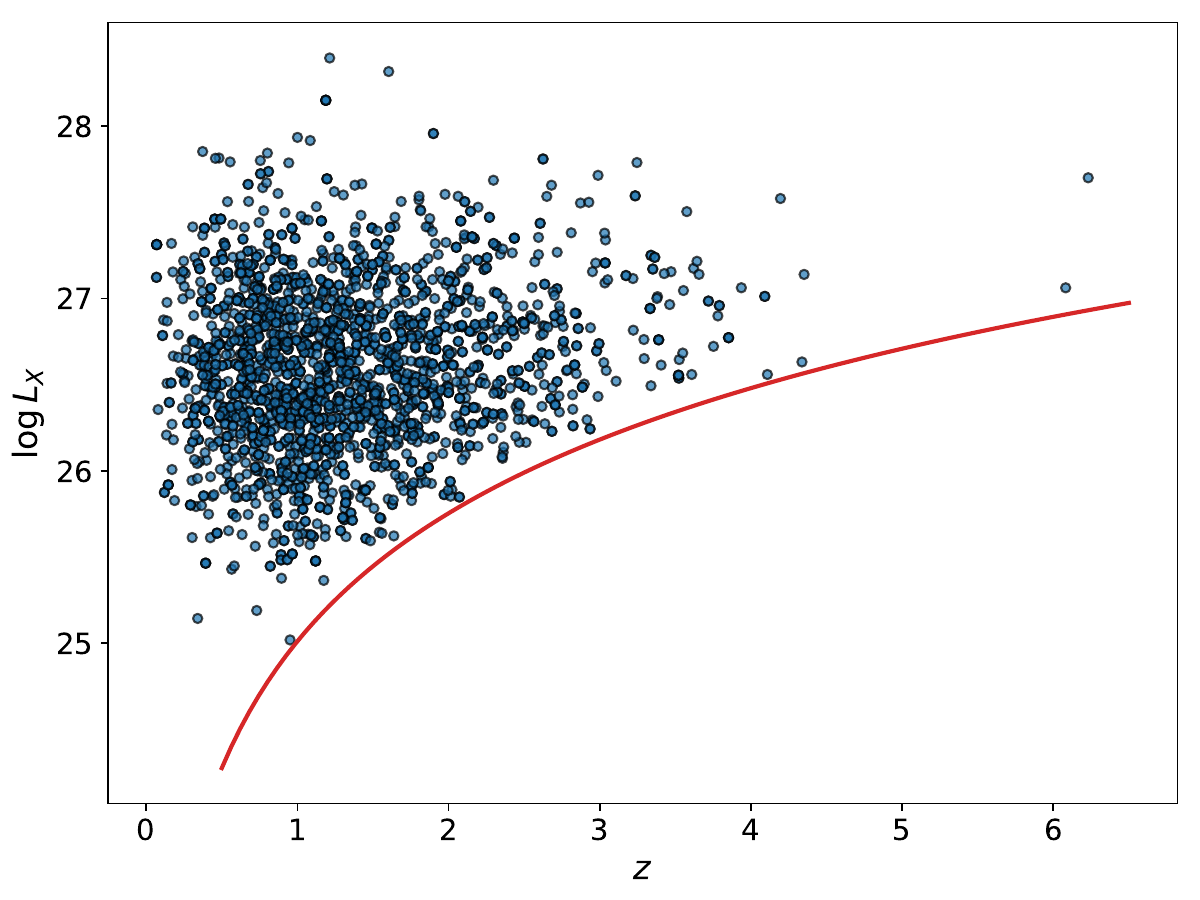}
\caption{\label{Fig2}  Simulated  $\log L_\mathrm{UV}$ and $\log L_\mathrm{X}$ versus redshift of quasars via the simulation method described in \Sec{Sec3}.}
\end{figure}

\begin{figure}[htbp]
\centering
\includegraphics[width=0.45\textwidth]{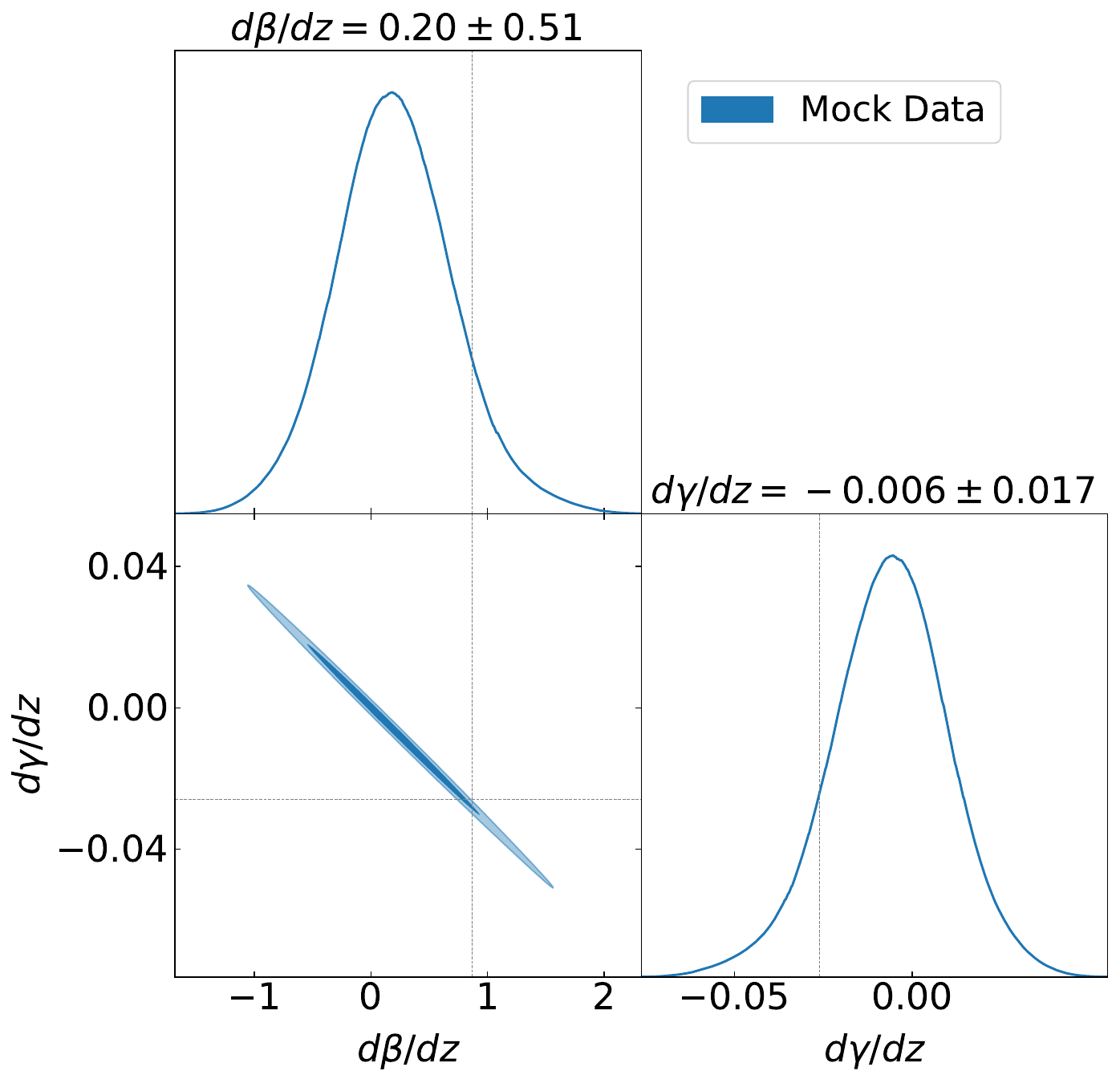}
\caption{\label{Fig3} Distribution of 1000 slope coefficients ($\frac{d\beta}{dz}$ and $\frac{d\gamma}{dz}$) calculated from the simulated quasars, with the dashed lines marking the mean values from the 2421 real quasars. }
\end{figure}

Now we can generate 2421 simulated quasars following the above simulation process, which are shown in Fig.~\ref{Fig2}.  Dividing the simulated data into four groups with the almost same number of data in each group, we can estimate the values of $\beta$ and $\gamma$ for each group, and then derive the $\beta-\langle z \rangle$ and $\gamma-\langle z \rangle$ relations, as  was done in the previous section.
By repeating this process 1000 times, we can derive 1000 slope  pairs ($\frac{d\beta}{dz}$ and $\frac{d\gamma}{dz}$) from  linear fits of the coefficients. The distribution of $\frac{d\beta}{dz}$ and $\frac{d\gamma}{dz}$ is presented as a contour plot in Fig.~\ref{Fig3}, in which the results from the real data are plotted with  dashed lines. One can see that the values of  $\frac{d\beta}{dz}$ and $\frac{d\gamma}{dz}$ from the real data deviate from those from the mock data at more than $1\sigma$ CL. The simulated data clearly do not support   redshift evolution of  the $L_\mathrm{X}$-$L_\mathrm{UV}$ relation since both $\frac{d\beta}{dz}$ and $\frac{d\gamma}{dz}$ are well consistent  with zero within $1\sigma$ CL. 

To further investigate the relation between the selection effect and the redshift evolutions of $\beta$ and $\gamma$, we introduce a redshift-dependent correction term $(1+z)^k$ to adjust the luminosity~\cite{Dainotti2022}, e.g. $L_\mathrm{corrected}=L_\mathrm{observed}/(1+z)^k$. We then apply the Efron-Petrosian (EP) method~\cite{Efron1992} to determine the value of the parameter $k$.
This analysis is performed using a publicly available Mathematica code, {\it  Selection biases and redshift evolution in relation to cosmology~\footnote{\href{https://notebookarchive.org/2023-05-8b2lbrh}{https://notebookarchive.org/2023-05-8b2lbrh}}}. 
If the values of $k$ derived from real data are significantly larger than those from mock data, it indicates that the selection effect alone cannot fully account for the redshift evolutions of $\beta$ and $\gamma$.
We obtain $k_\mathrm{UV}=3.632^{+0.069}_{-0.068}$ and $k_\mathrm{X}=2.689\pm0.053$ for real data, whereas for the mock data, we find  $k_\mathrm{UV}=0.106^{+0.097}_{-0.130}$ and $k_\mathrm{X}=0.460\pm0.087$. 
Clearly, the values of $k_\mathrm{UV}$ and $k_\mathrm{X}$ obtained from real data are  substantially larger than those from mock data, meaning that the evolutions of $\beta$ and $\gamma$ with redshift derived from real data cannot be attributed solely to the selection effects. This result is compatible with what is shown in Fig.~\ref{Fig3}

\section{Redshift-evolutionary  $L_\mathrm{X}$-$L_\mathrm{UV}$ relations}\label{Sec4}

\begin{figure}[htbp]
	\centering
	\includegraphics[width=0.46\textwidth]{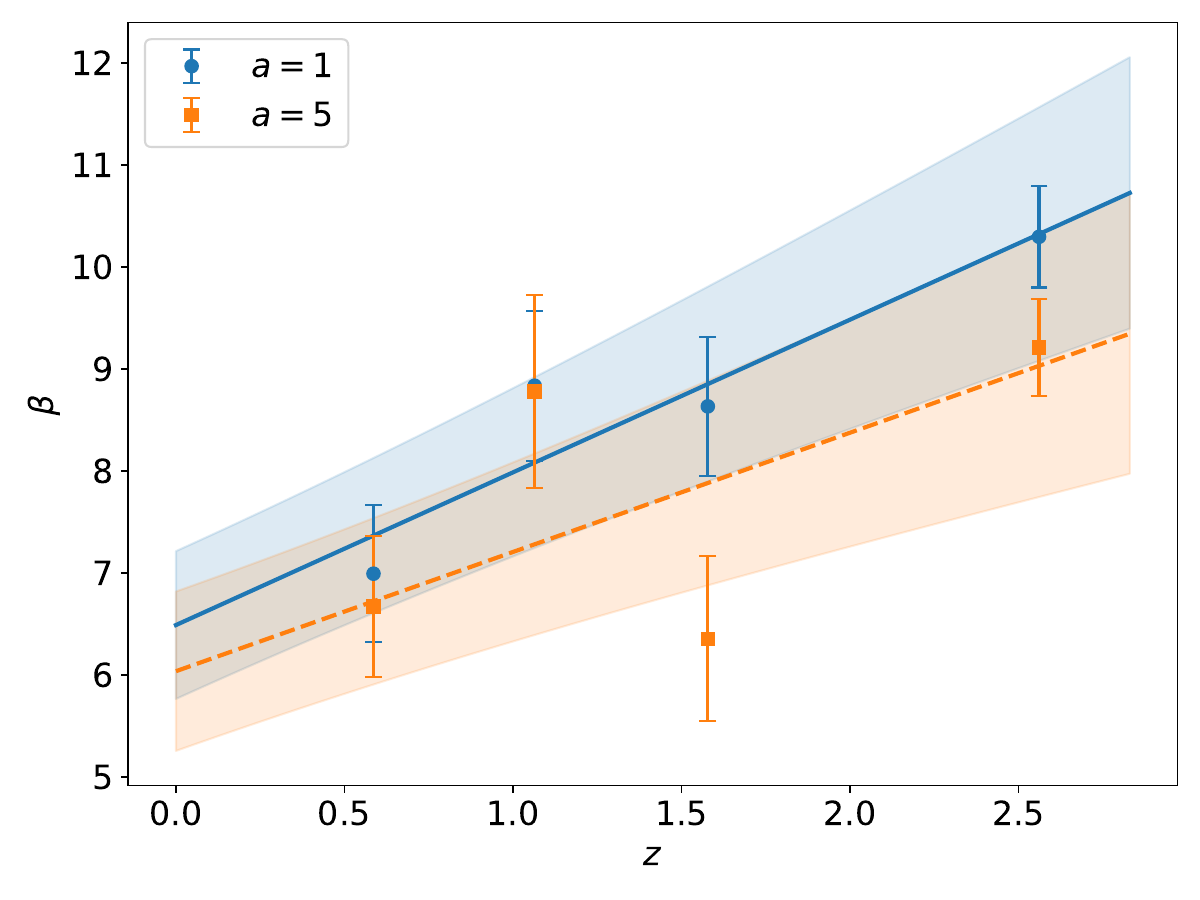}
	\includegraphics[width=0.47\textwidth]{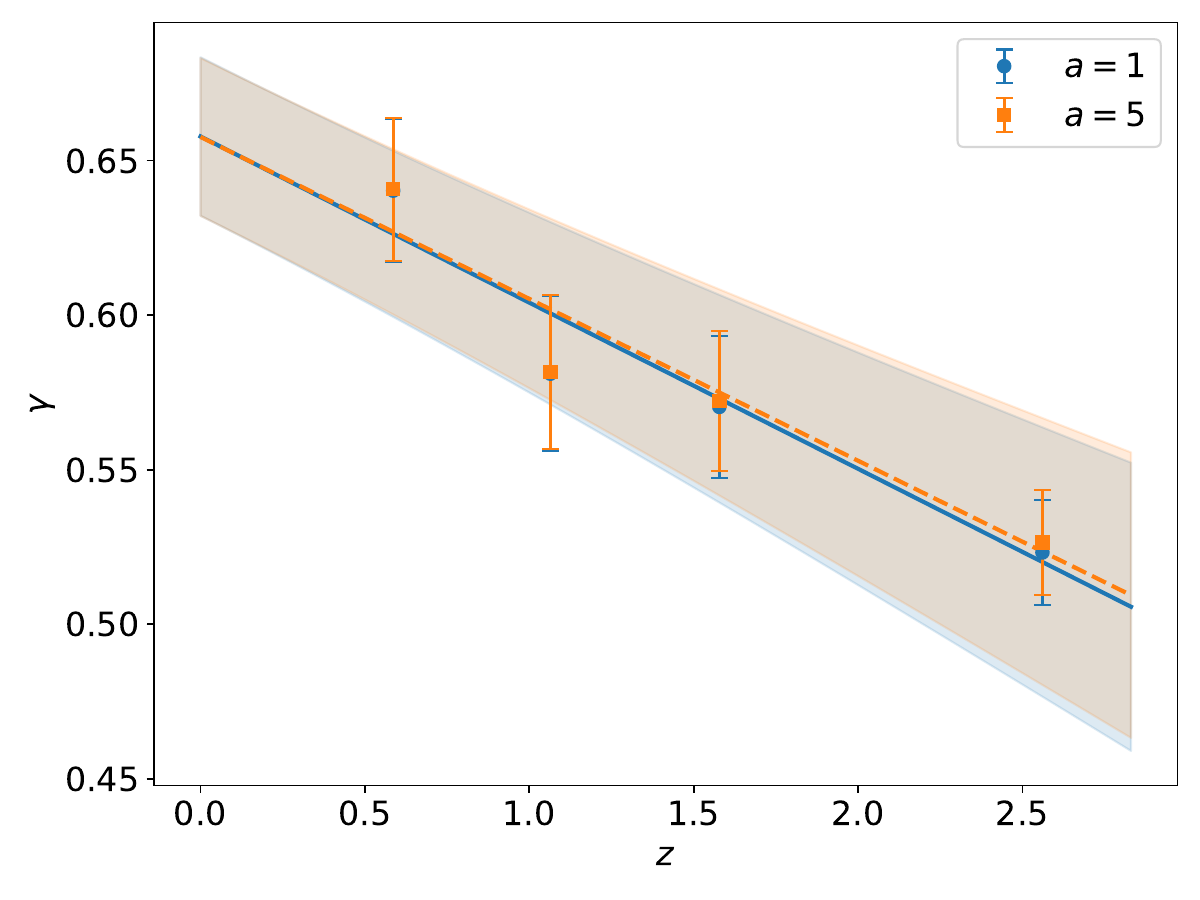}
	\includegraphics[width=0.47\textwidth]{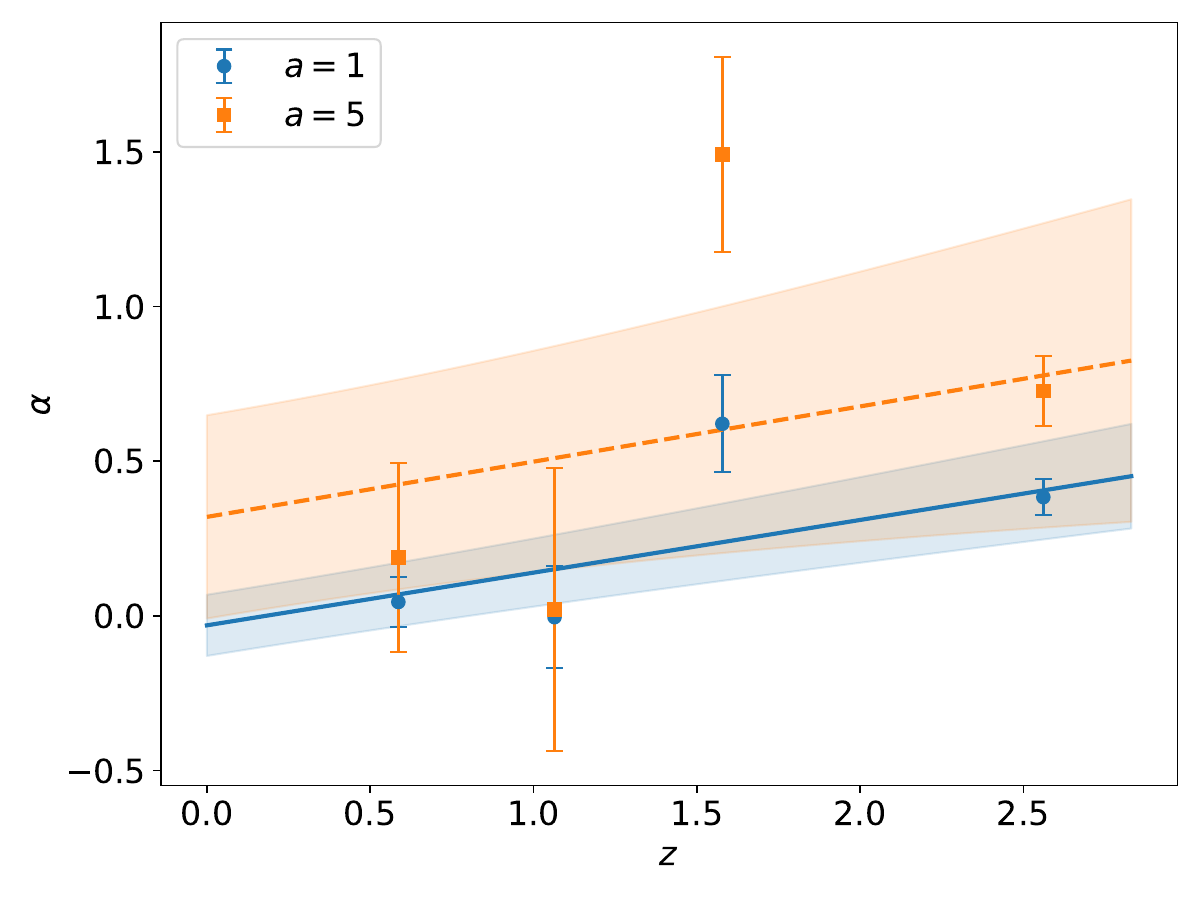}
	\caption{\label{Fig4} Values of $\beta$, $\gamma$, and $\alpha$ against the mean redshift of quasars based on the redshift evolutionary $L_\mathrm{X}$-$L_\mathrm{UV}$ relation (\Eq{eq.5}). Each data point with 1$\sigma$ error bar represents a group of quasars (1, 2, 3, and 4). The blue and orange points represent the different choices of $a$. The lines with shaded regions represent the linear fit with 1$\sigma$ uncertainty. }
\end{figure}

Now we consider two redshift evolutionary $L_\mathrm{X}$-$L_\mathrm{UV}$ relations, which can be expressed in the unified form: 
\begin{eqnarray}\label{eq.5}
	\log L_\mathrm{X}=\beta+\gamma \log L_\mathrm{UV}+\alpha \ln (a+z)
\end{eqnarray}
with $a=1$ and $a=5$, respectively. Here, $\alpha$ is a new parameter charactering the redshift-dependent property of the relation, and $\alpha=0$ corresponds to the case of the standard relation. When $a=1$, this relation is obtained by  assuming that the luminosities of quasars are corrected by a redshift-dependent function $(1 + z) ^\alpha$ as in~\cite{Dainotti2022}.  Substituting the corrected luminosities: $L_\mathrm{corrected: UV}=L_\mathrm{UV}/(1+z)^{k_\mathrm{UV}}$ and $L_\mathrm{corrected: X}=L_\mathrm{X}/(1+z)^{k_\mathrm{X}}$ into Eq.~(\ref{eq.1}) gives the relation shown in Eq.~(\ref{eq.5}) with $a=1$. When  $a=5$ the relation is constructed  using the Gaussian copula, as in~\cite{Wang2022}.   Copula,  a powerful tool, is used to describe the correlations among  $L_\mathrm{UV}$, $L_\mathrm{X}$ and $z$. Assuming that both $\log L_\mathrm{UV}$ and $\log L_\mathrm{X} $ follow  Gaussian distributions and the redshift of quasars also follows  a Gaussian distribution in the $z_*$ space with $z_*=\ln (a+z)$,  we construct the relation given by Eq.~(\ref{eq.5}). Wang et al.  found that choosing $a = 5$ provides an optimal fit after comparing several different values of $a$~\cite{Wang2022}.

We now check whether $\beta$, $\gamma$ and $\alpha$ evolve with redshift. 
Using the observed data of quasars, we calculate their values for each group, and show the corresponding results in Fig.~\ref{Fig4}. We find that the relation coefficients vary noticeably with redshift even when the redshift evolutionary relations are considered. The evolutionary trends of $\beta$ and $\gamma$  are almost identical to  those shown in Fig.~\ref{Fig1} where the standard $L_\mathrm{X}$-$L_\mathrm{UV}$ relation is considered. Thus, we can conclude that the redshift evolutionary $L_\mathrm{X}$-$L_\mathrm{UV}$ relations constructed in ~\cite{Dainotti2022,Wang2022} are not effective in
eliminating the redshift evolution of the relation coefficients. 

\section{A new $L_\mathrm{X}$-$L_\mathrm{UV}$ relation}\label{Sec5}

\begin{figure}[tbp]
 	\centering
 	\includegraphics[width=0.48\textwidth]{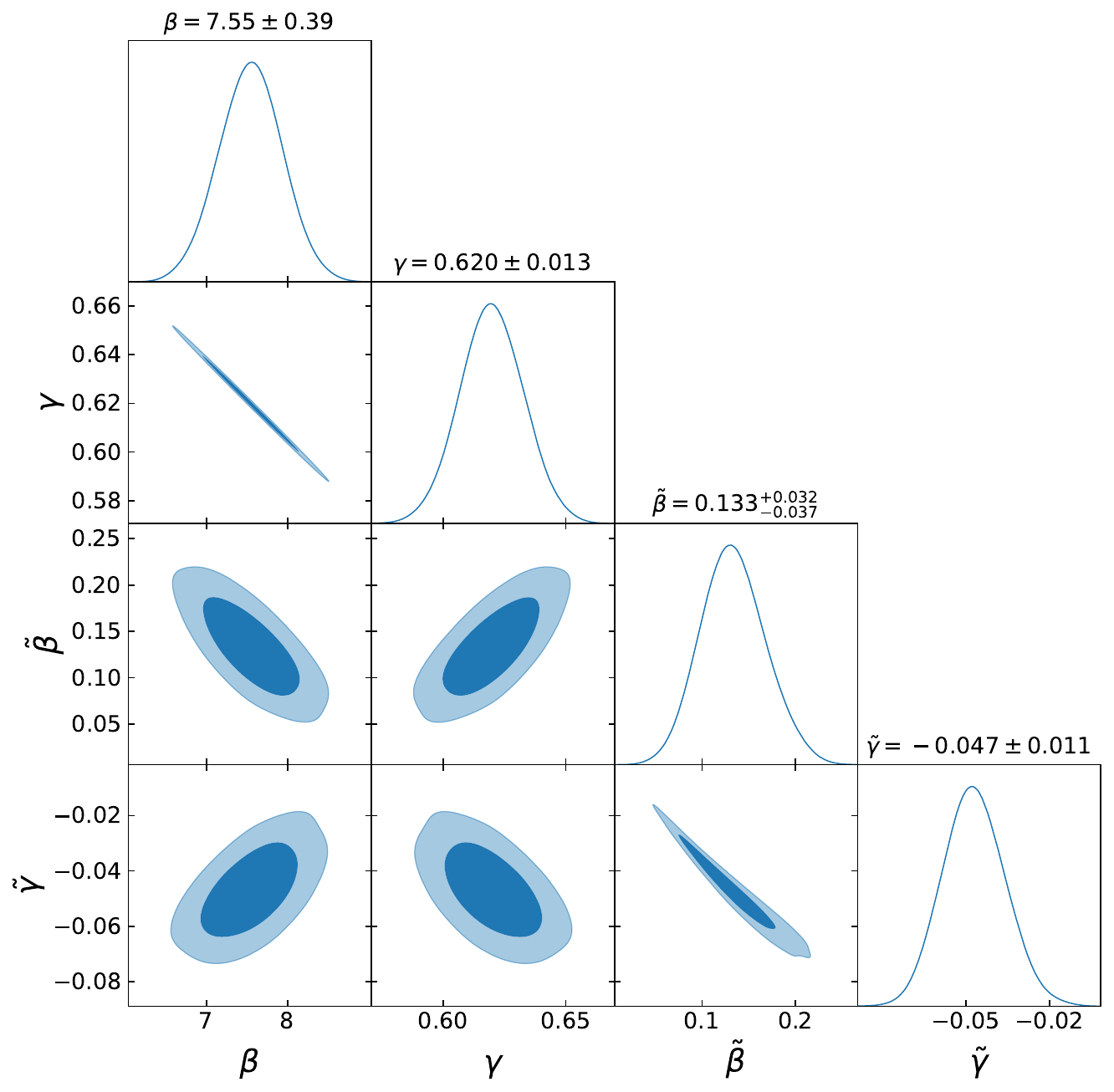}
 	\caption{\label{Fig5} { Constraints on coefficients ($\beta$, $\gamma$, $\tilde{\beta}$ and $\tilde{\gamma}$) of the new $L_\mathrm{X}$-$L_\mathrm{UV}$ relation (\Eq{eq.6}) from 2421 quasar data in the framework of the $\Lambda$CDM model with $\Omega_{m0}=0.3$}. The title in each subplot shows the  mean value with 1$\sigma$ uncertainty.}
 \end{figure}

We have established that the standard and two redshift evolutionary $L_\mathrm{X}$-$L_\mathrm{UV}$ relations evolve with redshift. In this section, we aim to to construct a new $L_\mathrm{X}$-$L_\mathrm{UV}$ relation  that eliminates the redshift evolution effect. As shown in Fig.~\ref{Fig1}, the coefficients $\beta$ and $\gamma$ are linearly dependent on redshift, motivating us to propose the following relation:
\begin{eqnarray}\label{eq.6}
	\log L_\mathrm{X}=\beta (1+\tilde{\beta}z)+\gamma(1+\tilde{\gamma}z) \log L_\mathrm{UV},
\end{eqnarray}
 where $\tilde{\beta}$ and $\tilde{\gamma}$ are new parameters introduced to account for any potential redshift evolution in the coefficients.  This  functional form is similar to  the one proposed in~\cite{Li2024}. We use the real data to constrain $\beta$,  $\gamma$, $\tilde{\beta}$ and $\tilde{\gamma}$ in the $\Lambda$CDM model with $\Omega_\mathrm{m0}=0.3$. The results  of our analysis are shown in Fig.~\ref{Fig5}. One can see that  the  coefficients $\tilde{\beta}$ and $\tilde{\gamma}$ exhibit significant  deviations  from zero, exceeding the 3$\sigma$ CL, indicating  that the quasar data strongly  support the redshift evolutionary relation. 
 
We then  proceed to  constrain the  values of the coefficients from four subsamples of quasars, which have almost the same data number, and  obtain 
\begin{eqnarray}
&& \beta=  (6.473\pm1.552)+(0.705\pm0.948)\times z,\\
&& \gamma= (0.657\pm0.054)+(-0.024\pm0.032)\times z
\end{eqnarray}
after using the linear fit. These coefficients exhibit weak dependence on redshift, as indicated by the slope values of
 ($\frac{d\beta}{dz}$ and $\frac{d\gamma}{dz}$),  both of which are consistent with zero at the $1\sigma$ CL.
Next, since the new $L_\mathrm{X}$-$L_\mathrm{UV}$ relation introduces two additional parameters compared to the standard one, we further constrain the values of $\beta$ and $\gamma$ by  fixing $\tilde{\beta}=0.133^{+0.032}_{-0.037}$ and $\tilde{\gamma}=-0.047\pm 0.011$, which are given by all quasars in the framework of the $\Lambda$CDM model. The resulting fits give us: 
\begin{eqnarray}
	&&\beta = (6.499\pm0.730) + (0.672\pm0.445) \times z,,\\
	&& \gamma=(0.656\pm0.025) + (-0.023\pm0.015) \times z.
\end{eqnarray}
Clearly,  $\beta$ and $\gamma$   evolve very weakly with redshift since the slopes $\frac{d\beta}{dz}$ and $\frac{d\gamma}{dz}$ are consistent with zero within $1.5\sigma$ CL.

These results show that the new   $L_\mathrm{X}$-$L_\mathrm{UV}$ relation effectively eliminates the redshift evolution effect on the relation coefficients. This new formulation provides a more stable and redshift-independent framework for modeling the  $L_\mathrm{X}$-$L_\mathrm{UV}$ relation in quasars.

\begin{figure}[htbp]
	\centering
	\includegraphics[width=0.46\textwidth]{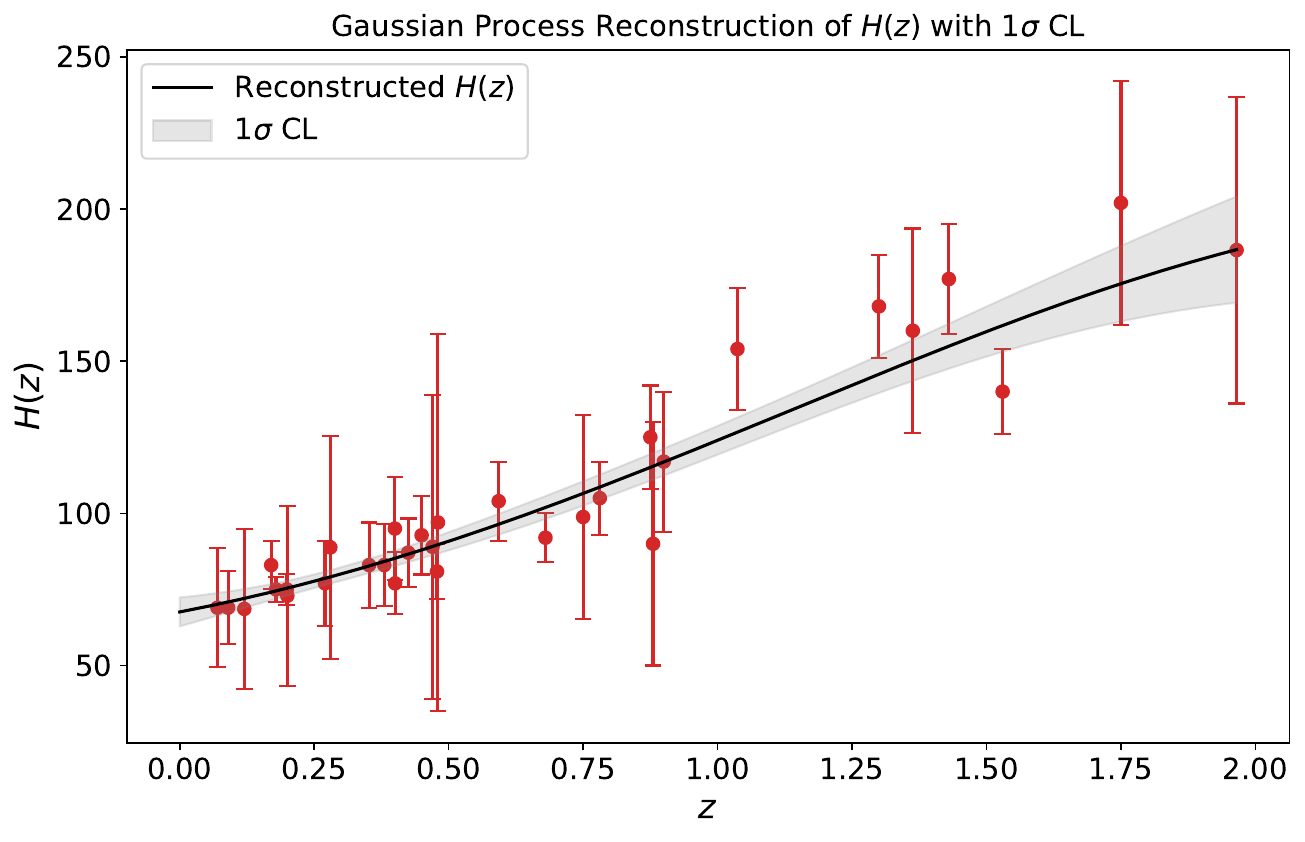}
	\includegraphics[width=0.47\textwidth]{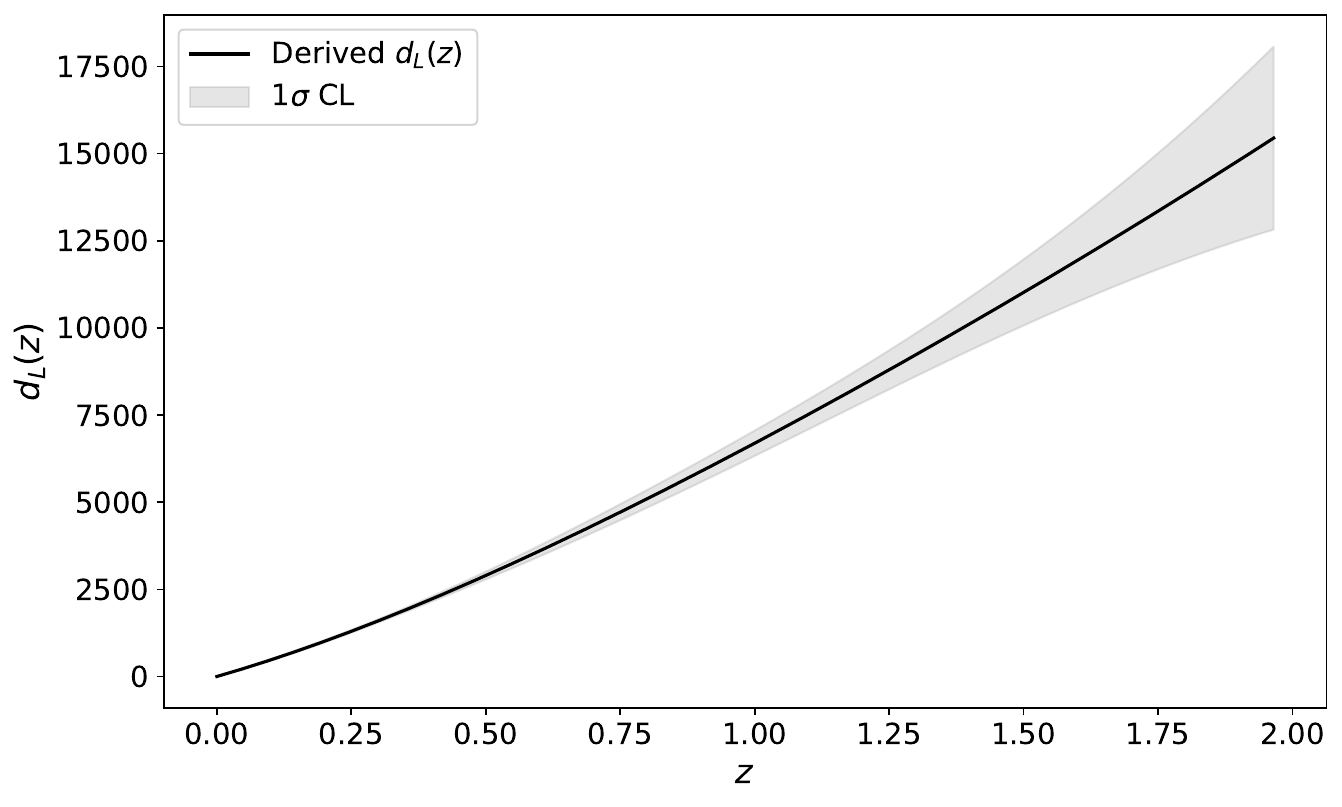}
	\caption{\label{Fig6} The reconstructed $H(z)$ from 32 Hubble parameter measurements, and the luminosity distance $d_L(z)$ derived from this $H(z)$. The shadows denote 1$\sigma$ uncertainty. }
\end{figure}

To investigate the effect of this new relation on the constraints of cosmological parameters, we fit the flat $\Lambda$CDM model and the flat $w$CDM model using quasars with the standard $L_\mathrm{X}$-$L_\mathrm{UV}$ relation and with the newly proposed relation  in Eq.~(\ref{eq.6}), respectively. We utilize  the Hubble parameter measurements to calibrate the luminosity relations of quasars.  The Gaussian Process (GP) method~\cite{Shafieloo2012, Marina2012,Cao2017b,Guo2025} is used to  reconstruct the $H(z)$ function from 32  observational data points  spanning a redshift range of $0.07 \leq z \leq 1.965$~\cite{Cao2022}. The reconstructed result is shown in the upper panel of Fig.~\ref{Fig6}.
From this reconstructed $H(z)$, we can derive  the luminosity distance $d_L(z)$ via the relation $d_L(z)=(1+z)\int_{0}^{z}1/H(z')\mathrm{d}z'$ as shown in the lower panel of Fig.~\ref{Fig6}.  Using the  luminosity distance $d_L(z)$, we can determine the values of the coefficients of the   $L_\mathrm{X}$-$L_\mathrm{UV}$ relations from 1892 quasars, which are within the low-redshift range ($z \leq 1.965$).  We obtain  $\beta=6.916\pm0.28$ and $\gamma=0.645\pm0.009$   at 1$\sigma$ CL for the standard relation, and $\beta=7.21\pm0.81$, $\gamma=0.631\pm0.027$, $\tilde{\beta}=0.187^{+0.091}_{-0.13}$ and $\tilde{\gamma}=-0.060^{+0.030}_{-0.035}$ at 1$\sigma$ CL  for the relation newly proposed in this paper. Extrapolating these results from the low-redshift quasars to  the full dataset of 2421 quasars, we then constrain the free parameters of the $\Lambda$CDM model and the  $w$CDM model from 2421 quasars.

\begin{figure*}[htbp]
	\centering
	\includegraphics[width=0.35\textwidth]{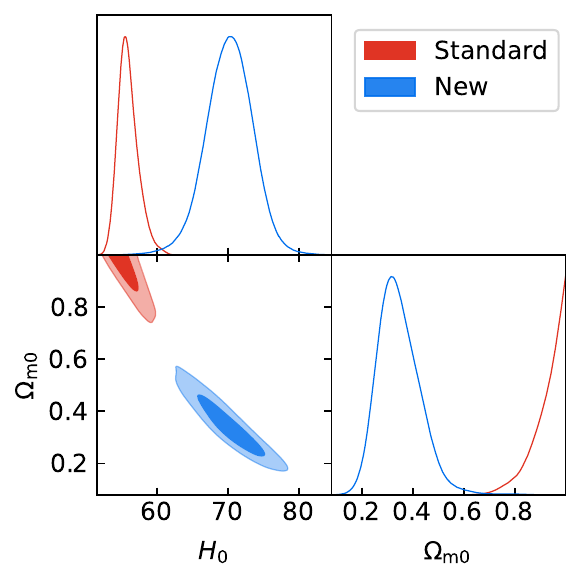}
	\includegraphics[width=0.45\textwidth]{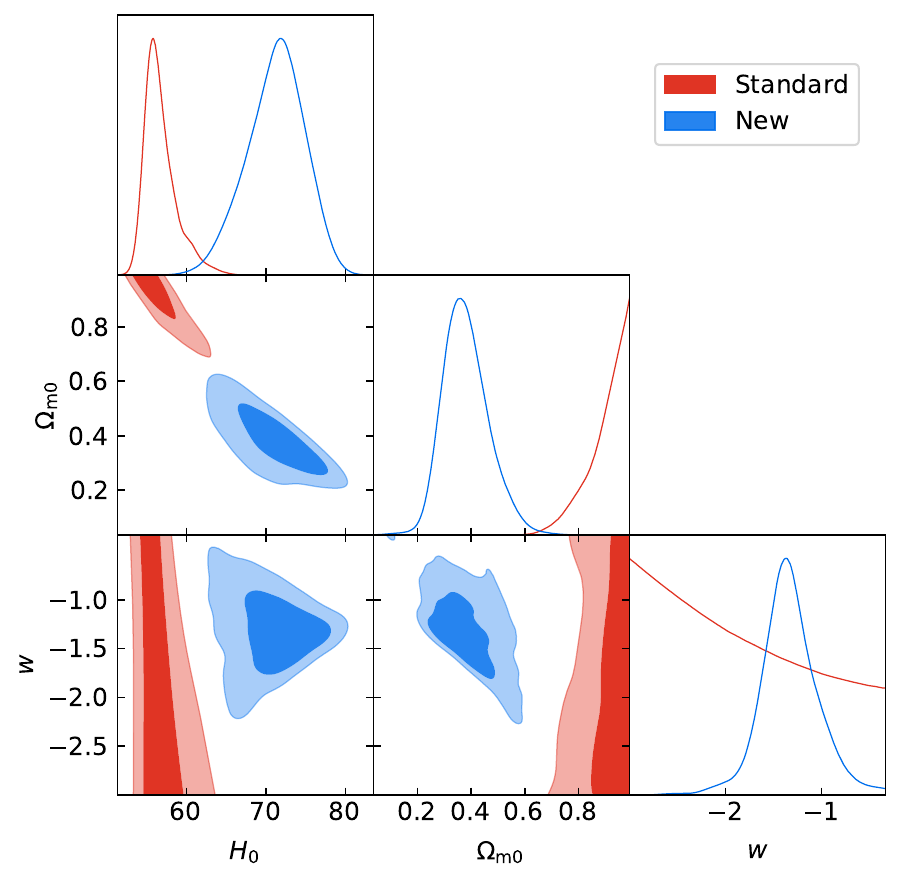}
	\caption{\label{Fig7} Constraints on the $\Lambda$CDM model and the $w$CDM model from full 2421 quasars standardized with the standard and new $L_\mathrm{X}$-$L_\mathrm{UV}$ relations, respectively. }
\end{figure*}

  The cosmological constraints are shown in Figure~\ref{Fig7}. For the $\Lambda$CDM model, we obtain $H_0=55.8^{+1.1}_{-1.6}~\kmsMpc$ and only a lower bound  $\Omega_{m0}>0.914$ using the standard  $L_\mathrm{X}$-$L_\mathrm{UV}$  relation. In contrast, using the new relation yields effective constraints, giving $H_0=70.2\pm3.3~\kmsMpc$ and $\Omega_{m0}=0.344^{+0.065}_{-0.093}$. The resulting $H_0$ value is consistent with measurements from Planck 2018 ($H_0=67.4\pm 0.5~\mathrm{km},\mathrm{s}^{-1},\mathrm{Mpc}^{-1}$)\cite{Planck} and from nearby type Ia supernovae calibrated with Cepheids ($H_0=73.04\pm 1.04\mathrm{km},\mathrm{s}^{-1},\mathrm{Mpc}^{-1}$)~\cite{Riess2022}.

 For the $w$CDM model, quasars  with the standard  relation provide weak constraints, yielding $H_0=56.7^{+1.1}_{-2.3}~\kmsMpc$, $\Omega_{m0}>0.883$, and $w<-1.45$. By employing the new relation, we achieve significantly improved constraints: $H_0=71.2^{+4.0}_{-3.2}~\kmsMpc$, $\Omega_{m0}=0.380^{+0.068}_{-0.095}$, and $w=-1.33^{+0.25}_{-0.29}$. 
These results align with those from the Planck 2018 CMB data within 2$\sigma$ CL~\cite{Planck}.
 Our findings clearly show that quasars standardized by the proposed relation can yield robust and effective cosmological constraints, unlike those standardized by the conventional relation.

 \section{Conclusions}\label{Sec6}
 Quasars play a crucial role as cosmological probes, and constructing accurate luminosity relations for them is essential for their use in cosmology.  In this paper,  we have thoroughly tested the redshift variation of the widely used $L_\mathrm{X}$-$L_\mathrm{UV}$ relation in quasars.  Our analysis reveals a strong linear dependence of the relation coefficients on redshift. Specifically, quasars at higher redshifts favor a larger $\beta$ and a smaller $\gamma$ compared to those at lower redshifts. These results are consistent with previous findings~\cite{Wang2022,Li2024}, where quasars were divided into high- and low-redshift subsamples.
 Importantly, this correlation is not due to the selection effect. 
 For two three-dimensional and redshift-evolving $L_\mathrm{X}$-$L_\mathrm{UV}$ relations, we observed  that the relation coefficients continue to evolve with redshift. This indicates that the redshift dependent terms in these evolutionary relations are not sufficient to eliminate the effect of redshift evolution in the coefficients. Finally, we  proposed a new $L_\mathrm{X}$-$L_\mathrm{UV}$ relation that contains four coefficients and found that, in this new relation,  the  relation coefficients evolve very weakly with redshift. This new relation suggests that the redshift evolution effect can be effectively minimized.  Using the Hubble parameter measurements to calibrate the newly proposed relation of quasars, we demonstrate that quasars can  effectively constrain on the $\Lambda$CDM and $w$CDM models. The resulting cosmological constraints are consistent with those derived from Planck CMB observations.  Our results indicate that  this new  $L_\mathrm{X}$-$L_\mathrm{UV}$ relation  allows quasars to be regarded as more reliable cosmological indicators. The elimination of redshift dependence in the relation could provide a more reliable framework for using quasars in cosmological applications.  
 
\begin{acknowledgments}
This work was supported by the National Natural Science Foundation of China (No.~12275080 and No.~12075084), and the Innovative Research Group of Hunan Province (Grant No.~2024JJ1006).
\end{acknowledgments}


\begin{thebibliography}{dummy}

\bibitem{Mortlock:2011}
    D. J. Mortlock,  S. J. Warren, B. P. Venemans, {\it et al.}, Nature, {\bf 474}: 616 (2011)
   
 \bibitem{Banados:2017}
  E. Banados, B.P. Venemans, C. Mazzucchelli, {\it et al.}, Nature, {\bf 553}:  473 (2018)
  
\bibitem{Lyke2020}
 B. W. Lyke,   A. N. Higley, J. N. McLane, {\it et al.},   Astrophys. J. Suppl., {\bf 250}: 8 (2020)
 
 \bibitem{Wang2021}
 F. Wang, J. Yang, X. Fan, {\it et al.},   Astrophys. J. Lett. {\bf 907}: L1 (2021)
 
 \bibitem{Zheng2020}
 X. Zheng, K. Liao, M. Biesiada, {\it et al.},  Astrophys. J., {\bf 892}: 103 (2020)
 
 \bibitem{Baldwin1977}
  J. A. Baldwin,  Astrophys. J., {\bf 214}: 679-684 (1977)
  

\bibitem{Osmer1999}
  P. S. Osmer and J. C. Shields,  ASP Conf. Ser., {\bf 162}: 235 (1999)
  
 \bibitem{Wang2014}J. M. Wang, P. Du, C. Hu, {\it et al.},  Astrophys. J., {\bf793}: 108 (2014)
 
\bibitem{Franca2014}
F. L. Franca, S. Bianchi, G. Ponti, {\it et al.}, Astrophys. J. Lett., {\bf 787}: L12 (2014)

\bibitem{Watson2011}
D. Watson, K. D. Denney, M. Vestergaard, and T. M. Davis,  Astrophys. J., {\bf 740}:  L49 (2011)

\bibitem{Melia2014}
F. Melia, JCAP, {\bf 01}: 027 (2014)

\bibitem{Eser2015}
E. Kilerci Eser, M. Vestergaard, B. M. Peterson, K. D. Denney, and M. C. Bentz,  Astrophys. J., {\bf 801}: 8 (2015)

 \bibitem{Paragi1999}
 Z. Paragi, S. Frey, L. I. Gurvits, {\it et al.},  Astron. Astrophys.,  {\bf 344}:  51 (1999)
 
 \bibitem{Chen2003}
 G. Chen and B. Ratra,   Astrophys. J.,  {\bf 582}: 586 (2003)
 
\bibitem{Cao2017}
S. Cao, X. Zheng, M. Biesiada, J. Qi, Y. Chen, and Z. H. Zhu,  Astron. Astrophys., {\bf 606}: A15 (2017)

\bibitem{Cao2020a}
S. Cao, J. Ryan, and B. Ratra,   Mon. Not. Roy. Astron. Soc., {\bf 497}: 3191-3203 (2020)

\bibitem{Cao2020b}
 S. Cao, J. Ryan, N. Khadka, and B. Ratra,   Mon. Not. Roy. Astron. Soc., {\bf 501}: 1520-1538 (2021)
 
\bibitem{Ryan2019}
 J. Ryan, Y. Chen, and B. Ratra,  Mon. Not. Roy. Astron. Soc.,  {\bf 488}: 3844-3856 (2019)
 
 \bibitem{Tananbaum1979}
H. Tananbaum, Y. Avni, G. Branduardi, {\it et al.}, Astrophys. J. Lett., {\bf 234}: L9 (1979)

\bibitem{Zamorani1981}
G. Zamorani, J. P. Henry, T. Maccacaro, {\it et al.},   Astrophys. J., {\bf 245}: 357  (1981)

\bibitem{Risaliti2015}
G. Risaliti and E. Lusso,  Astrophys. J., {\bf 815}: 33 (2015)

\bibitem{Avni1986} Y. Avni and H. Tananbaum,  Astrophys. J., {\bf 305}: 83 (1986)

\bibitem{Risaliti2019}
  G. Risaliti and E. Lusso,  Nature Astron., {\bf 3}: 272-277 (2019)
  
 \bibitem{Lusso2016}
  E. Lusso and G. Risaliti,  Astrophys. J., {\bf 819}: 154 (2016)
  
\bibitem{Lusso2017}
E. Lusso and G. Risaliti,  Astron. Astrophys., {\bf 602}: A79  (2017)

\bibitem{Lusso2020} E. Lusso,  G. Risaliti, E. Nardini, {\it et al.},   Astron. Astrophys., {\bf 642}:  A150 (2020)

\bibitem{Lusso2019}
E. Lusso, E. Piedipalumbo, G. Risaliti, M. Paolillo, S. Bisogni, E. Nardini, and L. Amati,  Astron. Astrophys., {\bf 628}: L4 (2019)

\bibitem{Lian2021}
Y. Lian, S. Cao, M. Biesiada, Y. Chen, Y. Zhang, and W. Guo,  Mon. Not. Roy. Astron. Soc., {\bf 505}:  2111-2123 (2021)

\bibitem{Hu2022}
J. P. Hu and F. Y. Wang,  Astron. Astrophys., {\bf 661}:  A71 (2022)

\bibitem{Khadka2020a}
N. Khadka and B. Ratra,  Mon. Not. Roy. Astron. Soc., {\bf 492}: 4456-4468 (2020)

\bibitem{Khadka2020b}
N. Khadka and B. Ratra, Mon. Not. Roy. Astron. Soc., {\bf 497}:  263-278 (2020)

\bibitem{Khadka2021}N. Khadka and B. Ratra,  Mon. Not. Roy. Astron. Soc., {\bf 502}: 6140-6156 (2021)

\bibitem{Li2021}
 X. Li, R. E. Keeley, A. Shafieloo, X. Zheng, S. Cao, M. Biesiada, and Z. H. Zhu,  Mon. Not. Roy. Astron. Soc.,  {\bf 507}: 919-926 (2021)
 

\bibitem{Li2022} Z. Li, L. Huang, and J. Wang,   Mon. Not. Roy. Astron. Soc., {\bf 517}: 1901-1906 (2022)



\bibitem{Wang2022} B. Wang, Y. Liu, Z. Yuan, N. Liang, H. Yu, and P. Wu,  Astrophys. J., {\bf 940}: 174 (2022)


\bibitem{Li2024}
  X. Li, R.  E. Keeley, and A. Shafieloo,  arXiv: 2408.15547
  
 \bibitem{Dainotti2022}  M. G. Dainotti, G. Bardiacchi, A. L. Lenart {\it et al.},  Astrophys. J., {\bf 931}: 106 (2022)
 
\bibitem{Nelsen} R. B. Nelsen, {\it An Introduction to Copulas} (New York: Springer, 2006)
  
\bibitem{Wang2024} B. Wang, Y. Liu, H. Yu, and P. Wu,  Astrophys. J., {\bf 962}: 103 (2024)


\bibitem{Zhang2024}
H. Zhang, Y. Liu, H. Yu, Xi. Nong, N. Liang, and P. Wu,   Mon. Not. Roy. Astron. Soc., {\bf 530}: 4493-4500 (2024)

\bibitem{DAgostini}G. D'Agostini,  arXiv: physics/0511182

\bibitem{Li2007}
L. X. Li,   Mon. Not. Roy. Astron. Soc., {\bf 379}: L55-L59 (2007) 

\bibitem{Cramer} H. Cra\'{m}er,   Scand. Actuarial J., {\bf 1928}: 13-74 (1928)

\bibitem{Mises}
 R. von Mises, {\it Wahrscheinlichkeit Statistik und Wahrheit} (Springer Berlin Heidelberg, 1928)
 
 \bibitem{Efron1992}
 B. Efron and V. Petrosian,  Astrophys. J., {\bf 399}: 345 (1992)

 \bibitem{Shafieloo2012}
  A. Shafieloo, A. G. Kim, and E. V. Linder,   Phys. Rev. D, {\bf 85}: 
123530 (2012)

\bibitem{Marina2012}
 S. Marina, C. Chris and S. Mathew,   JCAP,  {\bf 06}: 036 (2012)
 
\bibitem{Cao2017b} S. Cao, M. Biesiada, J. Jackson, X. 
Zheng, Y. Zhao, and Z. H. Zhu,   JCAP, {\bf 02}: 012 (2017)

\bibitem{Guo2025}
W. Guo, Q. Wang, S. Cao, {\it et al.},   Astrophys. J. Lett., {\bf 978}: L33 (2025)

\bibitem{Cao2022} S. Cao and B. Ratra,  Mon. Not. Roy. Astron. Soc., {\bf 513}: 5686-5700 (2022)


\bibitem{Planck} N. Aghanim, et al. (Planck),  Astron. Astrophys., {\bf 641}: A6 (2020)

\bibitem{Riess2022}  A. G. Riess, W. Yuan, L. M. Macri, {\it et al.},   Astrophys. J. Lett., {\bf 934}: L7 (2022)  


\end{thebibliography}

\end{document}